\documentclass[twocolumn,aps,prl,superscriptaddress,longbibliography]{revtex4-2}
\usepackage[colorlinks=true, citecolor=blue, urlcolor=blue, linkcolor=red]{hyperref}
\renewcommand{\section}[1]{{\par\it #1.~~}\ignorespaces}
\usepackage{amsmath,amssymb,tikz,scalerel}
\usetikzlibrary{svg.path}
\definecolor{orcidlogocol}{HTML}{A6CE39}
\tikzset{orcidlogo/.pic={
		\fill[orcidlogocol] svg{M256,128c0,70.7-57.3,128-128,128C57.3,256,0,198.7,0,128C0,57.3,57.3,0,128,0C198.7,0,256,57.3,256,128z};
		\fill[white] svg{M86.3,186.2H70.9V79.1h15.4v48.4V186.2z}
		svg{M108.9,79.1h41.6c39.6,0,57,28.3,57,53.6c0,27.5-21.5,53.6-56.8,53.6h-41.8V79.1z M124.3,172.4h24.5c34.9,0,42.9-26.5,42.9-39.7c0-21.5-13.7-39.7-43.7-39.7h-23.7V172.4z}
		svg{M88.7,56.8c0,5.5-4.5,10.1-10.1,10.1c-5.6,0-10.1-4.6-10.1-10.1c0-5.6,4.5-10.1,10.1-10.1C84.2,46.7,88.7,51.3,88.7,56.8z};}}
\newcommand\orcid[1]{\href{https://orcid.org/#1}{\mbox{\scalerel*{\begin{tikzpicture}[yscale=-1,transform shape]\pic{orcidlogo};\end{tikzpicture}}{|}}}}

\begin{document}
\title{Breakdown of Non-Bloch Bulk–Boundary Correspondence and Emergent Topology in Floquet Non-Hermitian Systems}
\author{Hong Wu\orcid{0000-0003-3276-7823}}\email{Contact author: wuh@cqupt.edu.cn}
\affiliation{School of Electronic Science and Engineering, Chongqing University of Posts and Telecommunications, Chongqing 400065, China}
\affiliation{Chongqing Key Laboratory of Dedicated Quantum Computing and Quantum Artificial Intelligence, Chongqing 400065, China}
\affiliation{Institute for Advanced Sciences, Chongqing University of Posts and Telecommunications, Chongqing 400065, China}
\author{Xue-Min Yang}
\affiliation{School of Electronic Science and Engineering, Chongqing University of Posts and Telecommunications, Chongqing 400065, China}
\affiliation{Chongqing Key Laboratory of Dedicated Quantum Computing and Quantum Artificial Intelligence, Chongqing 400065, China}
\affiliation{Institute for Advanced Sciences, Chongqing University of Posts and Telecommunications, Chongqing 400065, China}
\author{Hui Liu \orcid{0009-0009-4988-9561}}\email{Contact author: hui.liu@fysik.su.se}
\affiliation{Department of Physics, Stockholm University, AlbaNova University Center, 106 91 Stockholm, Sweden}

\begin{abstract}
Topological edge states in gaps of non-Hermitian systems are robust due to topological protection. Using the non-Hermitian Floquet Su–Schrieffer–Heeger model, we show that this robustness can break down: edge states may be suppressed by infinitesimal perturbations that preserve sublattice symmetry. We identify this fragility to the instability of the quasienergy spectrum in finite-size systems, leading to a breakdown of the non-Bloch bulk–boundary correspondence defined on the generalized Brillouin zone. To resolve this, we establish a correspondence between the number of stable zero-mode singular states and the topologically protected edge states in the thermodynamic limit. Our results formulate a bulk–boundary correspondence for Floquet non-Hermitian systems, where topology arises intrinsically from the driven non-Hermitian systems, even without symmetries. Our results provide a promising new avenue for exploring novel non-Hermitian topological phases.

\end{abstract}
\maketitle
\section{Introduction}
Due to the potential applications in novel devices, topological insulators have been extensively studied \cite{RevModPhys.82.3045,RevModPhys.91.015005,RevModPhys.91.015006}. This state of matter has a gap in the bulk band but conducting states on its edge. Such a property can be topologically protected by the symmetries of system \cite{RevModPhys.88.035005}. According to the principle of bulk-boundary correspondence, the order parameter called topological invariants which count the number of edge states can be defined \cite{RevModPhys.88.035005}. This has built a research paradigm for topological phases.

In recent years, the study of topological insulators has been extended to non-Hermitian systems, with the discovery of unique topological phenomena attracting considerable interest \cite{PhysRevLett.121.026808,PhysRevX.8.031079,PhysRevX.9.041015,Ashida02072020,PhysRevA.99.052118,PhysRevLett.123.016805,RevModPhys.93.015005,PhysRevLett.129.093001,PhysRevLett.128.223903,PhysRevLett.133.236902,pv42-w9r9,PhysRevLett.134.156601,wanjura2025unifyingframeworknonhermitianhermitian,PhysRevLett.134.056601}. Compared to Hermitian systems, there are two classes of complex-energy gaps in non-Hermitian systems: the point and line gaps \cite{PhysRevX.8.031079,PhysRevX.9.041015}. This leads to a classification of non-Hermitian systems into point-gap and line-gap topological insulators \cite{PhysRevLett.121.086803,Denner_2021}. The nontrivial point gap topology gives rise to a phenomenon in which abundant bulk states are localized at the edges, called skin effect \cite{PhysRevLett.124.086801,PhysRevLett.116.133903,PhysRevLett.125.126402,PhysRevA.109.L061501,PhysRevB.111.115415,PhysRevLett.129.086601}. Therefore, the energy spectra under open boundary condition are different from the periodic boundary ones \cite{xiong2017doesbulkboundarycorrespondence,PhysRevLett.116.133903}. We cannot characterize the edge states by the topological properties of the bulk bands \cite{xiong2017doesbulkboundarycorrespondence,PhysRevLett.116.133903}. This shows the breakdown of the bulk-boundary correspondence in non-Hermitian systems. By  replacing the Brillouin zone with the generalized Brillouin zone \cite{PhysRevLett.121.086803}, non-Bloch Hamiltonian-based topological invariants can be used to describe the number of edge states \cite{PhysRevLett.123.066404,PhysRevLett.125.186802}. This is called the non-Bloch bulk-boundary correspondence.

On the other hand, the concept of  topological insulators has been extended to out-of-equilibrium systems, such as periodically driven systems \cite{PhysRevX.3.031005,PhysRevB.96.155118,PhysRevB.100.085138,PhysRevResearch.1.022003,Merboldt_2025,Braun_2024,zhu2024characterizinggeneralizedfloquettopological}. The topological phases in this kind of system are called Floquet topological phases \cite{PhysRevLett.110.016802}. Compared to their static counterparts, Floquet systems show many novel phases, such as unique $\pi/T$-mode topological edge state \cite{tqyx-qyj4,kqly-w5d3} and the anomalous Floquet topological insulator, which hosts robust chiral edge modes even when all of
 its bulk Floquet bands carry trivial Chern numbers \cite{PhysRevX.3.031005,Maczewsky_2017}. These phases expand the family of topological states.

The merger of non-Hermitian and Floquet systems resulted in Floquet non-Hermitian topological phases. \cite{Xiao_2020}. The bulk-boundary correspondence has also been studied in such phases \cite{PhysRevB.102.041119,PhysRevB.103.L041115,zhou2025topologicalcharacterizationphasetransitions,PhysRevB.109.035418,PhysRevB.101.045415,PhysRevB.111.195419,PhysRevB.111.115424}. However, under open boundary conditions, the quasienergy spectrum is highly sensitive to variations in system size or symmetry-preserving disorder. Consequently, the introduction of non-Bloch band theory and the generalized Brillouin zone still does not suffice to establish a complete topological description for Floquet non-Hermitian systems. Thus, a general theory to characterize non-Hermitian Floquet topological phases is still lacking. 

In this work, we present a systematic investigation of topological insulators in periodically driven non-Hermitian systems. Motivated by the quasienergy spectrum instability in finite-size systems, we establish a correspondence between the number of stable 0-mode singular states and the protected topological edge states at quasienergies in the thermodynamic limit. We develop a unified theoretical framework in both momentum and real space to characterize the number of 0-mode and $\pi/T$-mode edge states. Our study introduces the bulk-boundary correspondence for Floquet non-Hermitian systems. Unlike previous studies, topology in this study is an intrinsic property of the driven non-Hermitian system, emerging even in the absence of symmetries.

\begin{figure}[tbp]
\centering
\includegraphics[width=1\columnwidth]{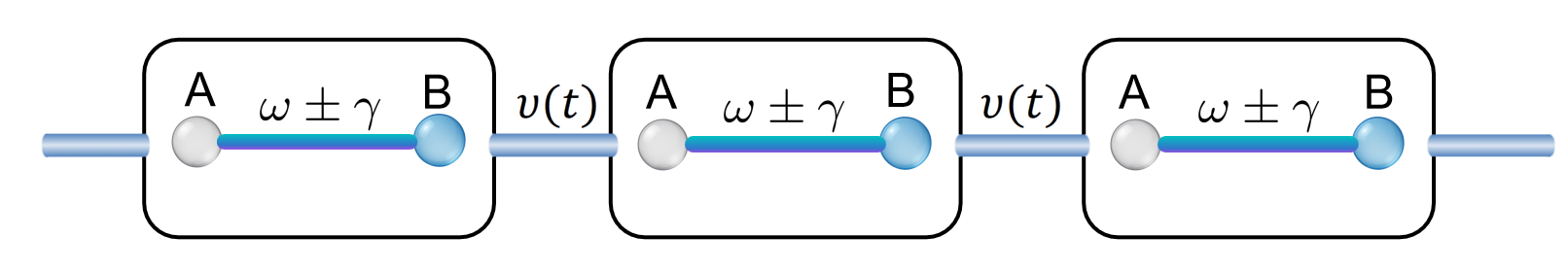}
 \caption{Schematics of the Su-Schrieffer-Heeger model on a chain.  The box indicates the unit cell.}
\label{tv11}
\end{figure}

\begin{figure}[tbp]
\centering
\includegraphics[width=1\columnwidth]{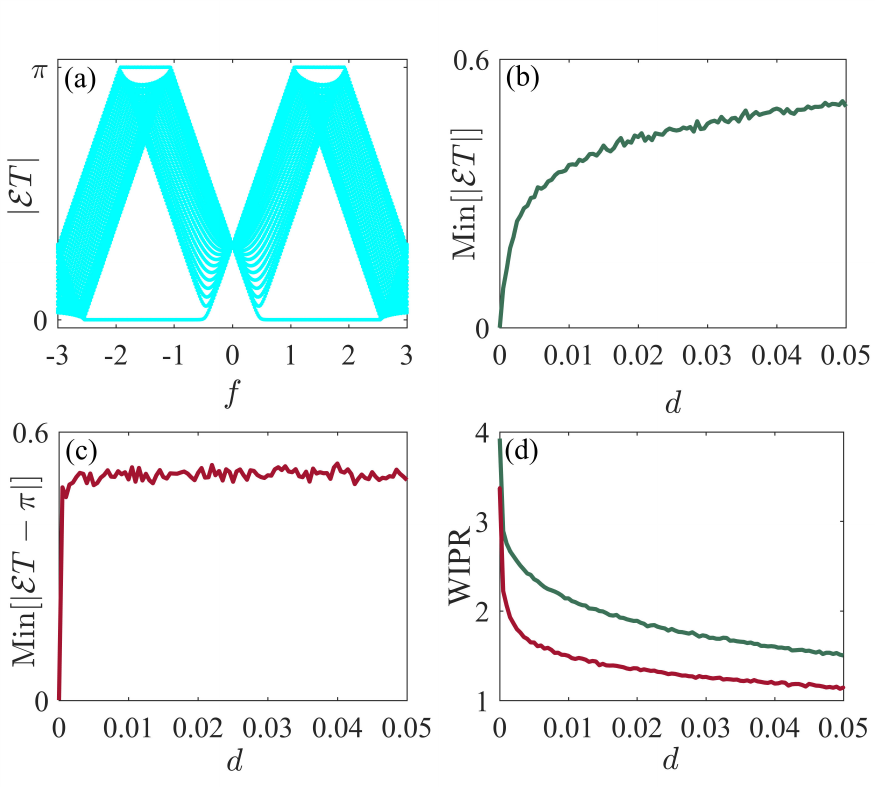}
 \caption{(a) Quasienergy spectra with the change of the driving amplitude under the open boundary conditions. The 0-mode (b) and (c) $\pi/T$-mode with the change of the disorder strength under the open boundary conditions. (d) disorder-averaged weighted inverse participation ratio. The results for green and crimson line are used $f=1$ and $f=1.5$, respectively. We use $w=1$, $\gamma=1.5$, $q=2$, $T_1=T_2=0.7$, and $N=25$. (b), (c), and (d) is obtained after 500 times average to the disorder.   
}
\label{tv22}
\end{figure}

\begin{figure*}
\centering
\includegraphics[width=1.76\columnwidth]{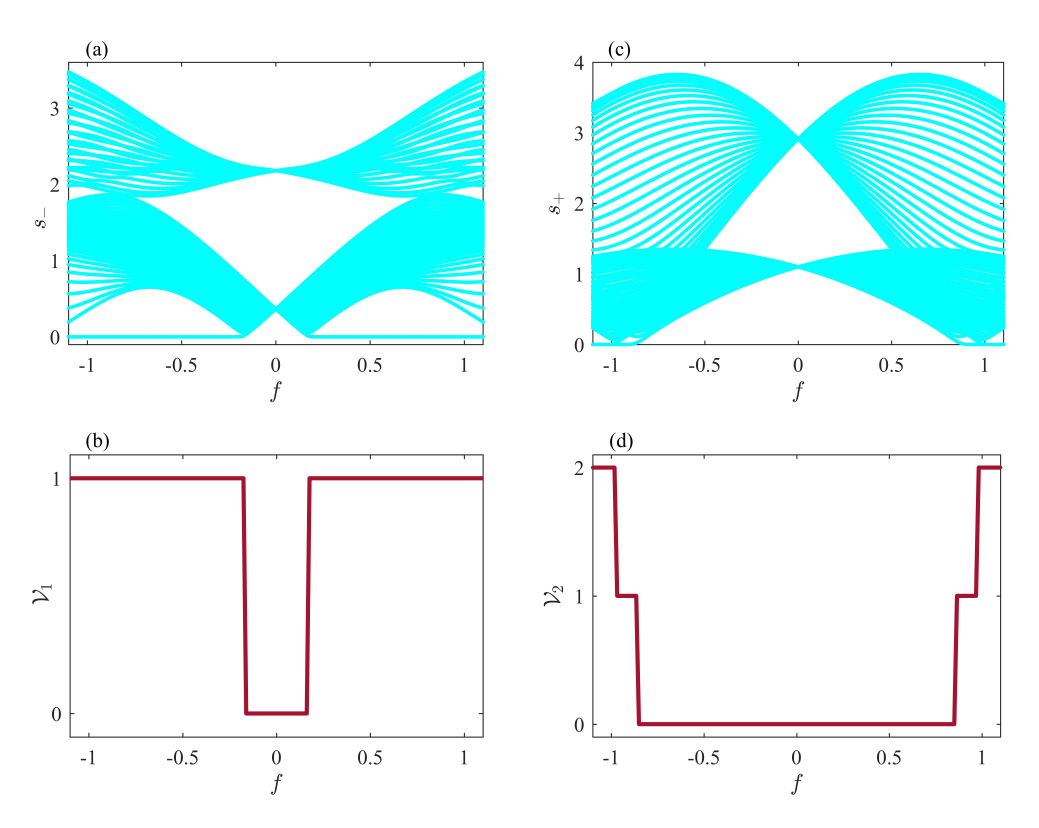}
 \caption{The singular spectra of $U(T)\pm I$ with the change of the driving amplitude under open boundary conditions in (a), (c) and corresponding winding number in (b), (d). We use $w=1$, $\gamma=1.5$, $q=2$, and $T_1=T_2=0.7$.  
}
\label{tv33}
\end{figure*}
\begin{figure}[tbp]
\centering
\includegraphics[width=1\columnwidth]{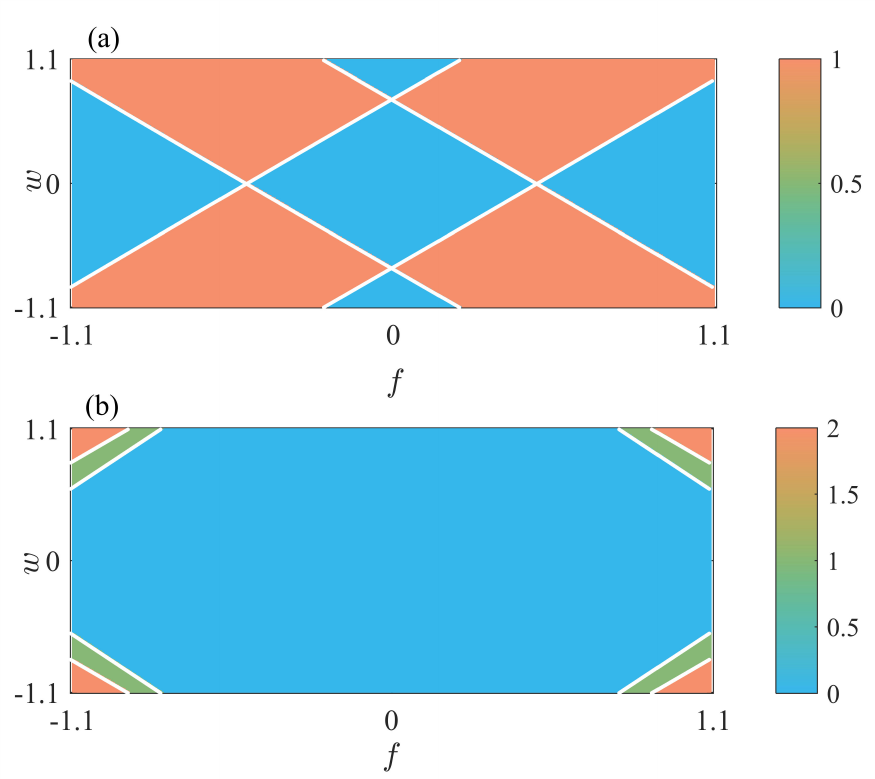}
 \caption{Phase diagram characterized by $\mathcal{V}_1$ (a) and $\mathcal{V}_2$ (b). The white lines are the phase boundaries that can be obtained from the band touching points of ${H}_{eff}$. We use $\gamma=1.5$, $q=2$, and $T_1=T_2=0.7$.}
\label{tv44}
\end{figure}

\section{The breakdown of non-Bloch bulk-boundary correspondence in Floquet system}Without loss of  generality, we demonstrate our result by the 1D non-Hermitian Su-Schrieffer-Heeger model \cite{PhysRevA.97.052115,PhysRevLett.121.086803}. Its Hamiltonian in real space is
\begin{eqnarray}
H=\sum_{l=1}^L[t_La^{\dag}_lb_l+t_Rb^{\dag}_la_l+v(a^{\dag}_lb_{l-1}+\text{h.c.})],~~~\label{Hmat}
\end{eqnarray}where $t_{L,R}=w\pm\frac{\gamma}{2}$, $a_l$ ($b_l$) are the annihilation operators on the sublattice A (B) of the $l$th unit cell, and $L$ is lattice length. In the momentum space, the Bloch Hamiltonian is
\begin{equation}
\mathcal{H}(k)=d_x\sigma_x+(d_y+i{\gamma}/{2})\sigma_y,\label{Hmt}
\end{equation}
where $d_x=w+v\cos k$ and $d_y=v\sin k$. This model has a sublattice
 symmetry $\sigma_z \mathcal{H}(k) \sigma_z=-\mathcal{H}(k)$, the $Z$  topological
 invariant can be defined as
 \begin{equation}
\mathcal{W}(\text{BZ})=\oint \frac{dk}{4\pi i}\text{tr}[\sigma_z\mathcal{H}^{-1}(k)\frac{d\mathcal{H}(k)}{dk}],
 \end{equation}
 where tr denotes the trace \cite{PhysRevLett.122.076801}. The number of edge states is $\mathcal{W}$. Due to the non-Hermitian skin effect, energy spectrum under the open boundary conditions has a dramatic difference from  the one under the periodic-boundary condition, this  can cause the breakdown of conventional bulk-boundary correspondence \cite{PhysRevLett.121.086803}. By introducing the generalized Brillouin zone (GBZ), non-Bloch bulk-boundary correspondence can be obtained \cite{PhysRevLett.121.086803}, $\mathcal{W}(\text{GBZ})$ counts the total number of zero modes.

Choosing the periodic driving as
\begin{equation}
v(t) =\begin{cases}f ,& t\in\lbrack mT, mT+T_1),\\q\,f,& t\in\lbrack mT+T_1, (m+1)T), \end{cases}~m\in \mathbb{Z}, \label{procotol}
\end{equation}
where $T$ is the driving period. See the schematic of this model in Fig. \ref{tv11}. This system does not have a well-defined energy spectrum.  According to Floquet theorem \cite{PhysRevLett.111.175301}, the one-period evolution operator ${U}(T)=\mathbb{T}e^{-i\int_{0}^{T}{H}(t)dt}$ defines an effective Hamiltonian ${H}_\text{eff}\equiv {i\over T}\ln [{U}(T)]$ whose eigenvalues are called the quasienergies. From the eigenvalue equation $U({T})\lvert u_l \rangle=e^{-i\mathcal{E}_{l}T}\lvert u_l \rangle$, we conclude that quasienergy $\mathcal{E}_l$ is a phase factor, which is defined modulus $2\pi/T$ and takes values in the first quasienergy Brillouin zone [$-\pi/T$,$\pi/T$] \cite{PhysRevLett.113.236803}. Topological phase of periodically driven system are defined in such quasienergy spectrum \cite{PhysRevB.96.195303}. We now investigate the Floquet topological phases in our periodically driven non-Hermitian SSH model.  Fig. \ref{tv22}(a) shows the quasienergy spectrum. There are both 0-mode and $\pi/T$-mode edge states in this Floquet system. In order to investigate the robustness of these states to disorder, we add the perturbation $\Delta{H}=d(\alpha_{ij}a^{\dag}_ib_j+\beta_{ji}b^{\dag}_ja_i)$ to the Hamiltonian $H(t)$, where $\alpha_{ij}$ and $\beta_{ji}\in [-0.5,0.5]$ is the disorder with strength $d$. Here $\Gamma \Delta H \Gamma^{-1}=-\Delta H$, where $\Gamma$ is the chiral operator in real space. Fig. \ref{tv22}(b) and \ref{tv22}(c) show the quasienergy of edge states with the change of  disorder strength. It is found that the edge states can be suppressed by sufficiently small perturbations that maintain the sublattice symmetry. There is no stability against small perturbations implying that these edge states don't have topological protection. To explain this phenomenon, we pay close attention to the skin effect in disorder system. Inspired by the definition of inverse participation ratio \cite{PhysRevLett.134.196302}, we construct a weighted inverse participation ratio:
\begin{equation}
\text{WIPR}=\frac{1}{2L}\sum_{n=1}^{2L}\sum_{x}|\psi_{n,x}|^4(x-L/2),
\end{equation}
where $x$ is the lattice site index. Here, \text{WIPR} can be used to describe the strength of non-Hermitian skin effect. Fig. \ref{tv22}(d) shows the weighted inverse participation ratio with the change of disorder strength. when $d$ is samll, \text{WIPR} has a rapid decay means the disruption of the skin effect. This simultaneously implies that the quasienergy spectrum of a finite-size system is highly sensitive to weak disorder; these boundary states are not protected by chiral symmetry. These reveal the failure of non-Bloch band theory in characterizing topology. we attribute this failure to the sensitive of boundary modes to finite-sized system, we next show in a themordynamic limit, we recover the non-Bloch bulk-boundary. 

\section{Restoration of bulk-boundary correspondence in Floquet non-Hermitian systems in momentum space }  The eigenspectrum in non-Hermitian system is unstable, so we propose a scheme to study the edge states of Floquet system by the spectrum of singular values. Consider the singular value decomposition $U(T)\pm I=U_{\pm}S_{\pm}V_{\pm}^{\dag}$ with $U_{\pm}$, $V_{\pm}$ unitary and $S_{\pm}$ diagonal and positive ($I$ is identity matrix). We denote the column vectors of $V_{\pm}$ ($U_{\pm}$) by $v_{n,\pm}$ ($u_{n,\pm}$) and the singular values on the diagonal of $S_{\pm}$ by $s_{n,\pm}$. We can obtain
\begin{eqnarray}
[U(T)\pm I]^{\dag}[U(T)\pm I]v_{n,\pm}=s^2_{n,\pm}v_{n,\pm}.
\end{eqnarray}
Then
\begin{equation}
[U(T)\pm I]v_{n,\pm}=s_{n,\pm}u_{n,\pm}.
\end{equation}
When $\lim_{L\rightarrow \infty}s_{-}=0$ ($\lim_{L\rightarrow \infty}s_{+}=0$), the quasienergies can have a 0-mode ($\pi/T$-mode) states in thermodynamic limit. Within this framework, the number of zero-mode singular values is directly linked to the number of topologically protected edge states in the quasienergy spectrum. Since the singular spectrum is highly robust to perturbations, as demonstrated in Fig. \ref{tv22}, it provides a useful approach for studying non-Hermitian topological phases in Floquet systems. $[U(T)\pm I]^{\dag}[U(T)\pm I]$ has the same bulk band under both open and periodic boundary conditions. Therefore, phase transition point can be given by $[U(T)\pm I]^{\dag}[U(T)\pm I]$ in momentum space. In momentum space, if the singular values are zero, the eigenvalues of $U(T)\pm I$ will also be zero. Therefore, the topological phase transition is associated with the closing of the quasienergy bands of Bloch effective Hamiltonian. The topological invariant that characterizes bulk topology can be defined as
\begin{eqnarray}
\mathcal{V}_1=\int_0^{2\pi}\frac{dk}{2\pi i}\partial_k\ln \det(U(T)-I),\label{tvq1}\\
\mathcal{V}_{2}=\int_0^{2\pi}\frac{dk}{2\pi i}\partial_k\ln \det(U(T)+I).
\label{tvq2}
\end{eqnarray}
According to Gohberg’s index theorem \cite{dingli}, $\mathcal{V}_1$ ($\mathcal{V}_2$) tells us directly about the
difference in the number of right and left 0 ($\pi/T$) eigenvalues of ${H}_{eff}$. Such topology is a generic property of periodically driven non-Hermitian systems which does not require the presence of any symmetries.

Fig. \ref{tv33} shows the singular spectra of $U(T)\pm I$ and corresponding topological number. The singular spectra can be well characterized by the two winding numbers $\mathcal{V}_1$ and $\mathcal{V}_2$. We can observe four typical regimes from this result: 
\\  
\text{(I)} There are one 0-mode edge state and two $\pi/T$-mode edge states for $0.97<\lvert f \lvert<1.1$. \\  
\text{(II)} There are one 0-mode edge state and one $\pi/T$-mode edge states for $0.87<\lvert f \lvert<0.97$ \\  
\text{(III)} One 0-mode edge state for $0.16<\lvert f \lvert<0.87$. \\
\text{(IV)} Topologically trivial phase when $\lvert f \lvert< 0.16$.\\
All edge states are localized on the left side. Compared to the static case, the Floquet system shows richer topological phases whose number of edge states is tunable. So far, we have established the Bulk-boundary correspondence in Floquet non-Hermitian systems. The method used to retrieve Bulk-boundary correspondence can be generalizable to other topological models and can provide a useful tool to study non-Hermitian topological phases in Floquet system.

To give a global picture of the non-Hermitian topological phases in our system, we plot in Fig. \ref{tv44} the phase diagram on the $f$-$w$ plane.  All phase boundaries match well with the band touching points of ${H}_{eff}$. Both $\mathcal{V}_1$ and $\mathcal{V}_2$ can be nonzero for some special parameters region, so the 0-gap and $\pi/T$-gap topology can coexist in a single system. This is a unique phase in the Floquet system. Besides, the number of edge states can be taken from 0 to 3.  This signifcantly expands the scope of the topological materials and enriches their
controllability.

\section{Restoration of bulk-boundary correspondence in Floquet non-Hermitian systems in real space} In finite-size systems with periodic boundary conditions, the eigenvalues of $U(T)\pm I$ may become unstable in the presence of disorder that breaks translational invariance. Therefore, the winding numbers Eq. \eqref{tvq1} and \eqref{tvq2} in real space are no longer valid. Here, we propose a more general scheme to retrieve the bulk-boundary correspondence. We consider 
\begin{equation}
\tilde{H}_{\pm}={
\left[ \begin{array}{cccc}
0 & U(T)\pm I & \\
U^{\dag}(T)\pm I& 0 & \\
\end{array}.
\right ]}.
\end{equation}
The eigenvalues of $\tilde{H}$ are given by $s_{n,\pm}$ and $-s_{n,\pm}$, where $s_{n,\pm}$ represents the singular values of $U(T)\pm I$ . The winding number of $\tilde{H}$ can be used as a topological number of ${H}_{eff}$ \cite{PhysRevB.103.224208}. \begin{eqnarray}
\mathcal{V'}_1=\frac{1}{4\pi i}\text{Tr} \ln (P_{-}^{A}{P_{-}^{B}}^{\dag}),\label{tvq3}\\
\mathcal{V'}_{2}=\frac{1}{4\pi i} \text{Tr} \ln (P_{+}^{A}{P_{+}^{B}}^{\dag}),
\label{tvq4}
\end{eqnarray}
where $P_{\pm}^{A}$ and $P_{\pm}^{B}$ are $P_{\pm}^{S}=U_{S,\pm}^{\dag}PU_{S,\pm}$, for $S=A,B$. $P_{ll'}=\delta_{ll'}e^{-i2\pi l/L}$. $U_{S,\pm}$ can be given by singular value decomposition 
\begin{equation}
U(T)\pm I=U_{A,\pm} s_{\pm} U^{\dag}_{B,\pm},
\end{equation}
where the diagonal elements $s_{\pm}$ is the singular values of $U(T)\pm I$. Here, the total number of right and left $0$-mode ($\pi/T$-mode) edge states in quasienergies is $\mathcal{V'}_1$ ($\mathcal{V'}_2$). The method presented here can also yield the results for the topological number in Fig. \ref{tv33}. So far, a more general description of Floquet non-Hermitian topological phases have been established in this work. 

\section{Possible experimental realization}
In recent years, Floquet non-Hermitian topological phases have been observed in open systems, such as photonics \cite{PhysRevLett.132.063804,PhysRevLett.133.073803} and quantum walk \cite{PhysRevLett.133.070801}. our Floquet model can also be simulated on open quantum system \cite{PhysRevLett.132.210402,huang2025complexfrequencyfingerprintbasic}.  To see the topology of open systems, we consider the equation of motion for the mean value of the single-particle operator  
 \begin{equation}
i\frac{d\lvert \mathbf{c}(t) \rangle}{dt}=H_{nh}\lvert \mathbf{c}(t) \rangle, \label{Xe}
 \end{equation}
 where $H_{nh}$ is an effective non-Hermitian Hamiltonian, $\lvert c(t) \rangle=\{\text{Tr}[a_1\rho(t)],\text{Tr}[b_1\rho(t)],......\text{Tr}[a_L\rho(t)],\text{Tr}[b_L\rho(t)]\\ \}^{T}$ \cite{huang2025complexfrequencyfingerprintbasic}.The EQ. \eqref{Xe} is analogous to the Schrödinger equation. Based on these developments, we believe that our approach is experimentally feasible. The topological modes can be detected by Loschmidt echo. 

\section{Conclusion and outlook}
 In summary, we have investigated the topological phases in periodically
driven non-Hermitian systems. Taking the non-Hermitian Su-Schrieffer-Heeger model as an example, we explained the breakdown of the non-Bloch bulk-boundary correspondence, which stems from quasienergy spectrum instability in finite-size systems. To solve this, we establish a correspondence between the number of stable 0-mode singular states and the protected topological edge states at quasienergies in the thermodynamic limit. A general description is established to characterize the number of 0-mode and $\pi/T$-mode edge states. Such result has not been reported before. 

Our work elucidates the topological characteristics of Floquet non-Hermitian systems. These results hopefully promote further studies of both fundamental physics and potential applications of this field. Besides, our scheme is generalizable to multiband non-Hermitian models with different topology to obtain bulk-boundary correspondence. It supplies a useful way to explore non-Hermitian Floquet topological phases.

\section{Acknowledgments}This work is supported by National Natural Science Foundation (Grants No. 12405007 and NO. 12305011), Funds for Young Scientists of Chongqing Municipal Education Commission(Grant No. KJQN20240 and N0. KJQN202500619), Natural Science Foundation of Chongqing (Grant No. CSTB2025NSCQ-GPX1265 and No. CSTB2022NSCQ-MSX0316), and Chongqing Natural Science Foundation Project (Grant No. CSTB2025NSCQ-LZX0142)

\bibliography{references}
\end{document}